# FROM CRYSTAL COLOR SYMMETRY TO QUANTUM SPACETIME

Martin Bojowald (Institute for Gravitation and the Cosmos, The Pennsylvania State University)
Avadh Saxena (Theoretical Division, Los Alamos National Laboratory)
Correspondence e-mail: bojowald@gravity.psu.edu, avadh@lanl.gov

**This perspective article elucidates both the importance and the implications of relativistic spacetime crystals as well as the renormalized blended coordinates transformation. It alludes to possible applications in materials science, condensed matter physics and quantum gravity.**

More than one hundred years after the inception of relativistic physics, the concept of time remains incompletely understood. Relativity provides means to perform calculations of geometrical properties of space-time, such as distances or curvature, and to interpret them in terms of physical observations, for instance as time dilation or gravitational effects. However, an intuitive understanding of space-time is complicated, not so much because it is 4-dimensional (which, after all, can be evaded by visualizing 2-dimensional cross-sections) but mainly because its geometry does not obey Euclid's axioms even in the absence of curvature.

Through a well-defined and clever transformation (RBS: Renormalized Blended Spacetime), Venkatraman Gopalan [1] has demonstrated how the hyperbolic geometry in Minkowski spacetime can be mapped to a circular Euclidean geometry (Figure 1). In particular, Lorentzian boosts become Euclidean rotations which enables new frontiers of exploration in color symmetry and magnetic crystals. His idea of general relativistic spacetime crystals and how to obtain them is both powerful and broad although the notion of relativistic crystals and lattices in two dimensions has existed for a while [2].

The breadth of this new work is underlined by the present opinion piece, written by two co-authors with distinct yet connected areas of expertise.

**Euclidean spacetime crystals**

As we know well, there are space crystals but in recent years there has been significant activity in exploring (quantum) time crystals since the suggestion of Wilczek in 2012 [3]. Time crystals refer to a system in which the lowest-energy states evolve periodically in time at a frequency different from that of the driving impulse. If spatial periodicities also arise simultaneously, these would be termed space-time crystals (note the *hyphenation* to indicate that time is disconnected from space). Space-time crystals are now known to exist in trapped ions, ultracold atoms and spin systems [4]. Recently, micrometer-sized, magnon based space-time crystals were created at room temperature [5].

Thus, this work is a natural generalization to relativistic *spacetime* crystals (note now, the lack of hyphenation between space and time to indicate that they can mix), in which both spatial and temporal periodicities coexist simultaneously. Beyond the spacetime crystals, an important consequence of RBS construction is that one can map the RBS symmetries to those of Euclidean point and space groups including the magnetic (or Shubnikov) or color groups [6], which is both insightful and very useful. The *color* in this context is a trait of whether an event is along time-like or space-like directions from the origin (Figure 1). A new anti-symmetry has been introduced which swaps time-like and space-like directions, thus reversing the color of this trait.

As far as color groups are concerned, in 1929 Heinrich Heesch introduced the anti-identity operation or the notion of anti-symmetry to enumerate 122 magnetic point groups associated with 22 magnetic Bravais lattices in three dimensions [7]. There are 31 such magnetic point groups corresponding to 5 magnetic Bravais lattices in two dimensions. Much later Aleksei Shubnikov fully developed the concept of magnetic symmetry [6]. Daniel Litvin has enumerated color or magnetic groups as well as he has computerized the process of obtaining crystallographic properties of such groups [8]. Many aspects of magnetic groups have already been adapted by the Bilbao crystallographic library [9]. These listings of space groups could thus find direct isomorphisms to RBS spacetime crystals.

In addition to introducing the RBS coordinates, the work of Gopalan provides a foundational framework for many further generalizations. (i) Using the RBS symmetries, deriving quasi-one dimensional (Q1D), Q2D and Q3D color groups will be a natural extension and very useful. Some important examples include color frieze groups, rod groups, diperiodic or layer groups, etc. (ii) Extension to 4D usual as well as color point and space groups would be a *tour de force* in crystallography. (iii) A more challenging extension could be deriving the usual and color quasi-crystalline groups. A quasi-crystal can be viewed as a crystal in an appropriate higher dimension. Thus, lower-dimensional color quasi-crystals could be derived from higher-dimensional color crystals with the newly found procedure. (iv) A more ambitious extension would be to derive symmetries in curved spacetime (e.g. in general relativity with Riemann geometry). These accomplishments will further enrich the crystallographic databases such as the one at Bilbao, Spain [9].

Just like their spatial counterparts, the spacetime crystals can also be quasiperiodic in time [4]. In addition, akin to the higher spatial dimensions, one can construct spacetime crystals with more than one time dimensions depending on how the system is driven by more than one periodic driving force. These possibilities could usher in an entirely new class of metamaterials with exotic properties otherwise not available in nature, besides understanding the fundamental attributes of a number of dynamical systems.

An important application of color symmetry is in magnetic phase transitions wherein symmetry is intimately related to the order parameter to construct the relevant free energy in Landau theory, in particular for studying these transitions in a variety of magnetic and multiferroic materials.

The latter have a coupling of magnetism to electric polarization or elastic strain. Interestingly, elastic strain and gravity have analogies through elastic compatibility condition and the Bianchi identity [10]. Clearly, Gopalan's seminal work has opened a richness of possibilities.

**Quantum spacetime: Flipping the minus sign**

A hallmark of important new results is their potential applicability in various contexts, a distinguishing feature that applies to Gopalan's construction. The hyperbolic geometry suitable for spacetime, introduced by Hermann Minkowski [11], may be viewed as a modification of Euclid's geometry in which the time direction contributes by a negative squared term in the Pythagorean theorem, relating the side lengths of right-angled triangles. Although a single minus sign might seem innocuous, the counter-intuitive nature of this non-Euclidean geometry is well-documented by the existence of several paradoxes in special relativity. While they can formally be resolved by doing suitable calculations in Minkowski geometry, they leave a nagging aftertaste of something not being quite right.

The mathematical formalism, of course, always gives a unique answer because special and general relativity are not only completely consistent as theories, but they have also been tested successfully by a good number of independent observations, most recently by the direct detection of gravitational waves [12]. However, all these results and tests refer to relativity as a classical theory. The other fundamental theories we know, describing the electro-weak forces and the strong nuclear force as well as the elementary particles that are subject to these forces, are quantum theories. According to general relativity, the geometry of spacetime depends, via Einstein's equation, on the energy density and pressure of all the matter it contains. If matter is quantum and fluctuates and jumps and gets entangled and what not, these effects should be transferred to spacetime geometry by a complete, quantum version of Einstein's equation.

It is hard to imagine what a fluctuating and entangled geometry, or a superposition of different geometries, might look like, and all attempts made so far in the endeavor called "quantum gravity" have led to mathematics of almost unmanageable complexity. (See for instance [13].) Moreover, the more successful ones of various approaches to this problem, given by string theory [14] and causal dynamical triangulations [15], work only because their math assumes that, for some reason, it should be OK to flip the sign in Minkowski's version of the Pythagorean theorem back to the positive value it has in Euclid's geometry. More briefly, these theories describe quantum properties of 4-dimensional space, not of spacetime.

There are various reasons for enforcing this sign flip. For one, the original negative sign implies instability because a time component of momentum would imply a negative contribution to the kinetic energy, which could then be minimized all the way to negative infinity. Gravity is indeed inherently unstable because it is always attractive and can imply perpetual collapse. The negative sign is therefore correct from a physical perspective, but the problem is that our usual methods of quantum field theory have a hard time dealing with it. If one flips the sign, these methods can at least be used to explore possible implications of quantized gravitational dynamics, even if they do not model the correct geometry.

The tacit assumption is that a quantum theory of 4-dimensional space is, in practice, close enough to a quantum theory of spacetime. Unfortunately, so far it has not been possible to justify this assumption, or to specify what "close enough" should mean in the preceding sentence. Here, Gopalan's "Relativistic Spacetime Crystals" is of considerable interest because it introduces an elegant renormalization procedure, RBS, that completes previous attempts of relating the geometries of space and spacetime. An application to quantum spacetime remains to be done, but by providing a fresh starting point, the paper's results promise further progress in this direction.

**Funding information**

This work was supported in part by the National Science Foundation (MB) and in part by the U.S. Department of Energy (AS).

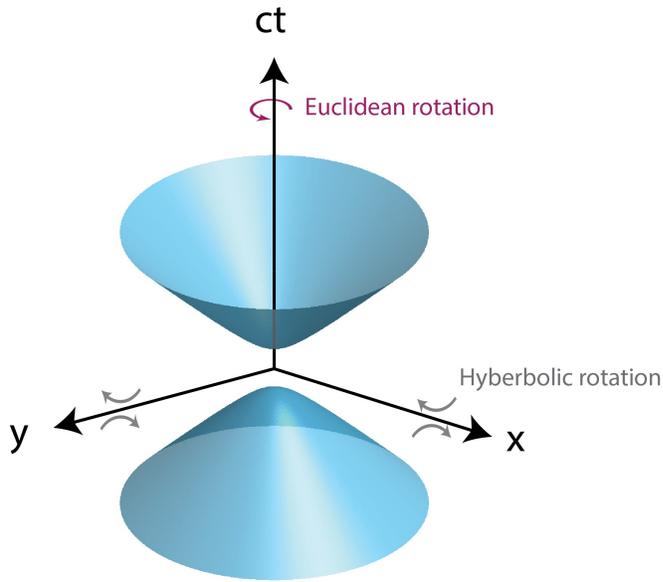

Symmetries of Minkowski spacetime

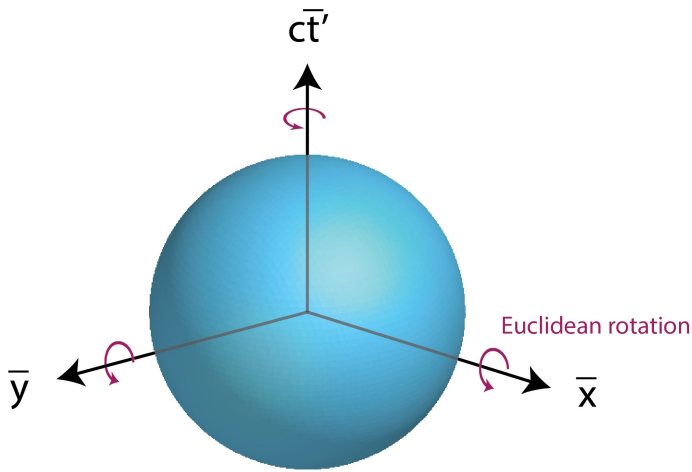

Symmetries of Renormalized Blended Spacetime (RBS)

**Figure 1:** Conventional 3D Minkowski spacetime (two space axes, x, and y, and one time axis, ct, depicted in the top panel) is hyperbolic. Events in the past and future at a fixed spacetime distance from the origin form two sheets of the hyperbola as shown. A new reformulation of special relativity by Gopalan, called the Renormalized Blended Spacetime (RBS), transforms the Minkowski hyperbolic sheets into a Euclidean sphere, (bottom panel). This topological transformation allows one to express the symmetries of the Minkowski spacetime, namely a combination of Euclidean and hyperbolic rotations, all as pure Euclidean rotations. The relativistic physics content in both formulations is equivalent. (Figure credit: Hari Padmanabhan, The Pennsylvania State University.)